\documentclass[aps,prl,twocolumn,showpacs,showkeys]{revtex4}
\usepackage{amssymb}
\usepackage{natbib}
\usepackage{amsmath}
\usepackage{amssymb}
\usepackage{amsfonts}
\usepackage{graphicx}
\usepackage{mathrsfs}
\usepackage{dcolumn}
\usepackage{bm}
\usepackage{color}

\definecolor{rot}{rgb}{0.75,0.05,0.25}
\definecolor{hellgrau}{gray}{0.5}
\definecolor{blau}{rgb}{0,0,0.7}

\begin{document}

\title{On a new definition of quantum entropy}
\author{Michele Campisi}
\email{campisi@unt.edu} \affiliation{Department of
Physics,University of North Texas Denton, TX 76203-1427, U.S.A.}
\date{\today }

\begin{abstract}
It is proved here that, as a consequence of the unitary quantum
evolution, the expectation value of a properly defined quantum
entropy operator (as opposed to the non-evolving von Neumann
entropy) can only increase during non adiabatic transformations
and remains constant during adiabatic ones. Thus Clausius
formulation of the second law is established as a theorem in
quantum mechanics, in a way that is equivalent to the previously
established formulation in terms of minimal work principle [A. E.
Allahverdyan and T. M. Nieuwenhuizen, Phys. Rev. E 71, 046107
(2005)]. The corresponding Quantum Mechanical Principle of Entropy
Increase is then illustrated with an exactly solvable example,
namely the driven harmonic oscillator. Attention is paid to both
microcanonical and canonical initial condition. The results are
compared to their classical counterparts.
\end{abstract}
\pacs{05.30.-d, 05.70.Ln} \keywords{Quantum thermodynamics, second
law, Clausius formulation, entropy increase} \maketitle

Allahverdyan and Niuwenhuizen \cite{Allahverdyan05} have recently
pointed out that the formulation of the second law of
thermodynamics in terms of Minimal Work Principle holds as a
quantum mechanical theorem. That is, the unitary law of quantum
mechanical evolution implies that \emph{the work $W$ done by an
external source on an isolated quantum system is larger than or
equal to the value $\widetilde{W}$ corresponding to the
quasi-static limit}. This is summarized as:
\begin{equation}\label{eq:W>0}
    W  \geq  \widetilde{W}
\end{equation}
The result in (\ref{eq:W>0}) is of great theoretical and practical
relevance as it shows in a clear way how thermodynamic laws
already exist at the microscopic quantum level. This is clearly
indicated also by the works of Jarzynski on non-equilibrium free
energy differences \cite{Jarz97}. These results and related ones
are now forming the basis of a new and exciting field of research
with a great potential impact on nano-science, known as
\emph{quantum thermodynamics} \cite{Allahverdyan04}.

In this Letter we will contribute to the field by showing
theoretically and with examples that the Clausius Principle of
Entropy Increase also holds as a theorem in quantum mechanics.
This is accomplished by introducing a new definition of quantum
entropy operator, which is alternative to von Neumann's one
\cite{Campisi08}:
\begin{equation}\label{eq:S=logN}
    \hat{\mathcal{S}}(t) \doteq \ln (\hat{\mathcal{N}}(t)+\hat{1}/2)
\end{equation}
where $\hat{\mathcal{N}}(t)$ is the time dependent \emph{quantum
number operator}:
\begin{equation}\label{}
    \hat{\mathcal{N}}(t) \doteq \sum_{k=0}^K k |k,t\rangle\langle k,t|
\end{equation}
The reasons for this choice will be clarified later.

For sake of clarity let us state Clusius Principle of Entropy
Increase here \cite{Uffink01}: \emph{For every non-quasi-static
process in a thermally isolated system which begins and ends in an
equilibrium state, the entropy of the final state is greater than
or equal to that of the initial state. For every quasi-static
process in a thermally isolated system the entropy of the final
state is equal to that of the initial state}. Thus, we are going
to prove and show with an exactly solvable example, that, for
every non adiabatic transformation the expectation value of the
quantum entropy operator of Eq. (\ref{eq:S=logN}) cannot decrease.
Further for every adiabatic transformation it remains constant.
Note that von-Neumann entropy ($-Tr \hat{\rho} \ln \hat{\rho}$)
cannot satisfy such prescription because it always remains
constant. Our result is slightly more general than Clausius
principle as we relax the requirement that the final state be an
equilibrium state.

Consider an isolated non-degenerate quantum system which is
initially in a state described by a density matrix
$\hat{\rho}(t_i) = \sum_{k=0}^K p_k |k,t_i\rangle \langle k,t_i|$
which is diagonal over the Hamiltonian eigenstates basis
$\{|k,t_i\rangle\}$. Let the density matrix eigenvalues be ordered
in a decreasing fashion:
\begin{equation}\label{eq:pm>pn}
    p_m \geq p_n \quad \text{if} \quad m < n
\end{equation}
At time $t_i$ a time dependent perturbation (not necessarily
small) is switched on so that the Hamiltonian of the system can be
expressed at any time $t>t_i$ as:
\begin{equation}\label{eq:spectum-H}
    \hat{H}(t) = \sum_{k=0}^K \varepsilon_k(t)|k,t\rangle \langle k,t|
\end{equation}
As time passes transitions will occur between the quantum states
according to the transition probabilities:
\begin{equation}\label{}
    |a_{kn}(t_f)|^2 = |\langle n,t_f|\hat{U}(t_i,t_f)|k,t_i\rangle  |^2
\end{equation}
Where $\hat{U}(t_i,t_f)$ is the unitary time evolution operator.
As a consequence the density matrix evolves to some
$\hat{\rho}(t_f)$. Eq. (\ref{eq:W>0}) has been proved under these
assumption with work defined as $W \doteq Tr \left[\hat{\rho}(t_f)
\hat{H}(t_f)- \hat{\rho}(t_i)\hat{H}(t_i)\right]$
\cite{Allahverdyan05}. Here we are interested in the change in the
expectation value of the entropy operator:
\begin{equation}\label{eq:Sf-si}
    S_f -S_i = Tr \left[\hat{\rho}(t_f) \hat{\mathcal{S}}(t_f) - \hat{\rho}(t_i)
    \hat{\mathcal{S}}(t_i)\right]
\end{equation}
Let us denote the probability that the system is found in state
$n$ at time $t_f$, by $p'_n$, where evidently $p'_n= \sum_{k=0}^K
p_k|a_{kn}(t_f)|^2$, then
\begin{equation}\label{eq:Sf-si}
    S_f -S_i = \sum_{n=0}^K (p'_n-p_n) \ln (n+1/2)
\end{equation}
Using the ``summation by parts'' rule \citep{Allahverdyan02},
Eq. (\ref{eq:Sf-si}) becomes
\begin{equation}\label{eq:DeltaS}
    S_f -S_i = \sum_{m=0}^K \ln \frac{m+\frac{3}{2}}{m+\frac{1}{2}} \sum_{n=0}^m (p_n-p'_n)
\end{equation}
Using the property of the coefficients $ |a_{kn}|^2$ of forming a
doubly stochastic matrix (i.e., $\sum_{k=0}^K |a_{kn}(t_f)|^2 =
\sum_{n=0}^K |a_{kn}(t_f)|^2 = 1$ and $|a_{kn}(t_f)|^2 \geq 0 $),
and the ordering of probabilities (\ref{eq:pm>pn}),
it can be shown \cite{Allahverdyan05,Campisi08} that $
\sum_{n=0}^m (p_n-p'_n) \geq 0$, which, by noting that $\ln
\frac{m+\frac{3}{2}}{m+\frac{1}{2}} > 0$ in Eq. (\ref{eq:DeltaS}),
implies
\begin{equation}\label{eq:Entropy-Increase}
    S_f \geq S_i.
\end{equation}
This holds as a consequence of the laws of quantum mechanics for
every time dependent perturbation acted on a non-degenerate
quantum mechanical system. As such it is an exact
\emph{non-equilibrium} result. In case the perturbation is
adiabatic, the quantum adiabatic theorem would ensure that no
transition will occur between states with different quantum number
\cite{Messiah62} so that $p'_n=p_n$ and consequently $S_f=S_i$.
Eq. (\ref{eq:Entropy-Increase}) is the \emph{Quantum Mechanical
Principle of Entropy Increase}. 
Note that, like Clausius Principle, Eq.
(\ref{eq:Entropy-Increase}) in no ways implies that the
expectation value of the entropy is monotonic increasing. Eq.
(\ref{eq:Entropy-Increase}) says \emph{only} that after time $t_i$
the expectation value of the entropy operator will never be less
than the initial value. This does not rule out the possibility
that for two times $t_1>t_2>t_i$ one might have $S_1<S_2$.

The reason for defining the entropy operator as in
Eq.(\ref{eq:S=logN}) can be understood by considering the
classical microcanonical entropy \cite{Gibbs02}:
\begin{eqnarray}
    S^{cl}(E,t) &=& \ln \Phi(E,t) \\
    \Phi(E,t) &\doteq& \int_{H(\mathbf{q},\mathbf{p};t)\leq E} \frac{d^{3N}\mathbf{q}
    d^{3N}\mathbf{p}}{h^{3N}} \label{eq:volumeEntropy}
\end{eqnarray}
that is the logarithm of the volume $\Phi$ of classical phase
space enclosed by the hyper-surface of energy $E$. Using the
semi-classical viewpoint \citep[pp. 23-24]{Landau5} the integral
in Eq. (\ref{eq:volumeEntropy}) counts the number of quantum
states not above a certain energy $\varepsilon_n = E$. Since the
levels are assumed to be non degenerate, this number is
$n+\frac{1}{2}$, where we set by convention that the vacuum state
counts as a half state. Thus one can construct the quantum version
of Eq. (\ref{eq:volumeEntropy}) and obtain Eq. (\ref{eq:S=logN}).
By repeating all the calculations reported above in the classical
case one can prove that also the classical expectation value of
the classical entropy cannot decrease \cite{Campisi08}. This
requires replacing the discrete quantity $n+1/2$ with the enclosed
volume $\Phi$, and the probability $p_n$ with the probability
density function $P(\Phi)$ of having one member of the ensemble
being on an orbit that encloses a volume $\Phi$. Accordingly all
sums over $n$ will be replaced with integrals over $d\Phi$. The
coefficients $|a_{kn}|^2$ have to be replaced by the ``classical
transition probability'' $A(\Phi, \Theta)$ that a representative
point which is on an orbit that encloses a volume $\Phi$ at time
$t_i$ will be found on an orbit
that at time $t_f$ encloses a volume $\Theta$. 
Thus, if the initial distribution is $P(\Phi)$, the final one is
$P'(\Theta)=\int d\Phi A(\Theta,\Phi)P(\Phi)$ \cite{Campisi08}.
Liouville's Theorem implies that the ``classical transition
probability'' $A(\Phi, \Theta)$ is doubly stochastic, i.e., $\int
A(\Theta,\Phi)d\Phi=\int A(\Theta,\Phi)d\Theta=1$
\cite{CampisiDiss}.

If one translates the quantum mechanical requirement of
non-degeneracy (only one state per energy eigenvalue) with the
classical requirement of ergodicity (only one trajectory per
energy level) \footnote{This is necessary for the classical
validity of Thomson formulation too. See A. E. Allahverdyan and T.
M. Nieuwenhuizen, Phys. Rev. E 75, 051124 (2007).}, and the
requirement in Eq. (\ref{eq:pm>pn}) with the requirement that
$P(\Phi)$ be monotonic decreasing, then one obtains
\cite{Campisi08} (compare with Eq. (\ref{eq:Sf-si})):
\begin{equation}\label{eq:DeltaScl}
    S^{cl}_f-S^{cl}_i=\int_0^{\infty}d\Phi(P'(\Phi)-P(\Phi)\ln\Phi \geq 0
\end{equation}

Let us illustrate these results with a practical example of an
exactly solvable time-dependent quantum system. Consider the
driven harmonic oscillator:
\begin{equation}\label{eq:omegaHO}
    H(x,p,t)=\frac{p^2}{2m}+\frac{m}{2}\omega^2 x^2 + f(t)x
\end{equation}
which can be solved analytically and is relevant in many
applications such as polyatomic molecules in varying external
fields, colliding polyatomic molecules and electrons in crystals
\cite{Feynman65}. For simplicity let us set $m=1$, $\omega=1$,
$t_i=0$, $t_f=T$, and consider the case when the perturbation is
\emph{cyclic} and switched on only during the time interval
$[0,T]$\cite{Feynman65}. Let us pick a representative point from
our initial ensemble, located at $x_0,p_0$ at time $t_i$, 
and draw in phase space the circular orbit
$H(x,p,0)=H(x_0,p_0,0)$. Let $r_0$ be the radius of such orbit,
then $\frac{r_0^2}{2}$ will be the energy $E_0$ of the system and
the enclosed volume $\Phi$ will be given by $\frac{2\pi}{h}
E_0=\frac{E_0}{\hbar}$. Let us then choose the units in such a way
that $\hbar = 1$. Then, in this specific problem, \emph{energy and
enclosed volume coincide}. Using Laplace transforms one can easily
solve the classical problem and obtain the energy $E$ at time $T$,
as: $E = E_0 + W +2\sqrt{E_0W}\cos(T-\varphi-\theta) $ or, using
the enclosed volume notation (i.e., $E_0=\Phi, E=\Theta$):
\begin{equation}\label{eq:Theta}
    \Theta = \Phi + W(T) + 2\sqrt{\Phi
    W(T)}\cos(T-\varphi-\theta(T))
\end{equation}
where 
\begin{eqnarray}
  2W(T) &=& |\beta(T)|^2, \quad \theta(T)=\arg \beta(T) \label{eq:2W(t)}\\
  \beta(T) & \doteq & \int_0^{T}dt f(t) e^{i\omega t}
\end{eqnarray}
Thus, depending on the initial phase $\varphi$, a particle of
initial enclosed volume $\Phi$ will have different final enclosed
volume $\Theta$ at time $T$. The probability of ending up
enclosing a volume $\Theta$ from a point that initially encloses a
volume $\Phi$, then is easily calculated as
$A(\Theta,\Phi)=1/\left( \pi
\left|\frac{d\Theta}{d\varphi}\right|\right)$. Using Eq.
(\ref{eq:Theta}) we get:
\begin{equation}\label{eq:A}
    A(\Theta,\Phi)=\frac{1}{\pi}\frac{1}{\sqrt{4\Phi W-(\Theta-\Phi -W)^2}}
\end{equation}
One can check that $A(\Theta,\Phi)$ is doubly stochastic. Using
Eq. (\ref{eq:A}) one finds the average microcanonical final
enclosed volume and its square deviation:
\begin{eqnarray}
    \left\langle \Theta \right\rangle_\Phi &=& \Phi + W \label{eq:cl-first}\\
    \left\langle \Theta^2 \right\rangle_\Phi - \left\langle \Theta
    \right\rangle_\Phi^2 & = &2\Phi W \label{eq:cl-second}
\end{eqnarray}
where the symbol $\left\langle \cdot \right\rangle_\Phi$ denotes
average over the initial microcanonical distribution of energy
$E_0=\Phi$. Note, from Eq. (\ref{eq:cl-first}), that $W$ is
actually the work done on the initial ensemble. The microcanonical
expectation of final entropy is most conveniently calculated
directly from Eq. (\ref{eq:Theta}) as \cite{table07}:
\begin{eqnarray}
\left\langle \ln\Theta \right\rangle_\Phi &= &\int_0^{2\pi}
\ln(\Phi + W + 2\sqrt{\Phi W}\cos\varphi)d\varphi \nonumber \\
& = & \ln [\max(\Phi,W)]  \label{eq:dScl-mc}
\end{eqnarray}
Note that trivially one has $\left\langle \ln\Theta
\right\rangle_\Phi \geq \ln \Phi$. That is, Clausius formulation
is satisfied in this case even for an initial microcanonical
distribution. According to Eq. (\ref{eq:2W(t)}), $W$ is definite
positive, so also the Minimal Work Principle (\ref{eq:W>0}), which
for a cyclic process reads $W \geq 0$ (Thomson formulation) is
satisfied. Let us calculate the entropy change resulting from an
initial canonical distribution:
\begin{equation}\label{}
P(\Phi)d\Phi=\beta e^{-\beta \Phi }d\Phi
\end{equation}
\begin{figure}
\includegraphics[width=7 cm]{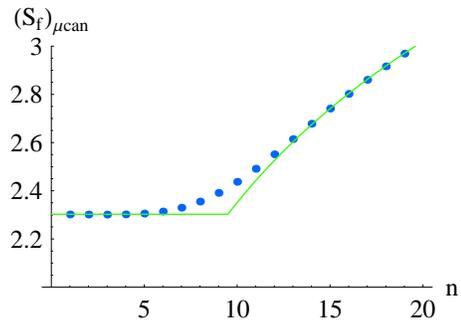}
  \caption{Microcanonical expectation of entropy as function of the initial quantum number $n$ for $W=10$. Solid line: classical case,
  Eq. (\ref{eq:dScl-mc}).
 Dots: quantum case, Eq. (\ref{eq:dSqu-mc}), the sum has been truncated at $m=1000$. The quantum phenomena smooth out the sharp angle in the classical graph.}
\label{fig:fig1}
\end{figure}
From Eq. (\ref{eq:DeltaScl}) we get:
\begin{eqnarray}
S^{cl}_f-S^{cl}_i &=& \beta\int_0^{\infty}  d\Phi e^{-\beta \Phi}
[\left\langle \ln\Theta \right\rangle_\Phi-\ln\Phi] \nonumber \\
&=&-\beta W \int_0^1 dx e^{-\beta W x}\ln x \label{eq:dScl-c}
\end{eqnarray}
which is evidently positive as expected.

Let us now move to the quantum mechanical case. The transition
probabilities $|a_{nm}|^2 $ are expressed in terms of Charlier
polynomials $C(m,n|W)$ \cite{Husimi53,Feynman65}:
\begin{eqnarray}\label{eq:Charlier}
    C(m,n|W)=\sum_l^{\min(m,n)}\frac{(-1)^l m! n!}{l!
(m-l)!(n-l)!W^l}\\
    |a_{nm}(T)|^2= \frac{e^{-W(T)}W^{m+n}(T)}{m!n!}[C(m,n|W(T))]^2
\end{eqnarray}
The average microcanonical quantum number and its square deviation
are analogous to the classical formulas \cite{Husimi53} (compare
with Eqs. (\ref{eq:cl-first},\ref{eq:cl-second})):
\begin{eqnarray}\label{}
    \left\langle m \right\rangle_n &=& n+W \label{eq:qu-first}\\
\left\langle m^2 \right\rangle_n - \left\langle m
\right\rangle_n^2 &=& 2(n+1/2)W
\end{eqnarray}
where the symbol $\left\langle \cdot \right\rangle_n$ denotes
average over the initial quantum mechanical microcanonical
distribution $\hat{\rho}_i = |n,0\rangle \langle n,0|$. The
average microcanonical quantum expectation of entropy reads
\begin{equation}\label{eq:dSqu-mc}
    \left\langle \ln (m+1/2) \right\rangle_n = \sum_m |a_{nm}(t)|^2 \ln (m+1/2)
\end{equation}
Unlike the average quantum number and square deviation this does
not correspond exactly to the classical expression. Figure
\ref{fig:fig1} shows the quantum and classical average
microcanonical expectation of entropy. Note that the quantum
effects smooth out the sharp angle that appears in the classical
case. Note also that Clausius principle is satisfied for the
microcanonical initial condition also in the quantum case, that is
$\left\langle \ln (m+1/2) \right\rangle_n \geq \ln (n+1/2)$. Since
any distribution is a superposition of microcanonical
distributions, then Clausius formulation is satisfied here for
\emph{any} initial distribution, without restriction to decreasing
ones (\ref{eq:pm>pn}). This is true for the Minimal Work Principle
as well (\ref{eq:qu-first}). Let us calculate the change in
quantum expectation of entropy resulting from an initial canonical
distribution:
\begin{equation}\label{}
    \hat{\rho}_i= (1-e^{-\beta})\sum_n e^{-\beta n} |n,0\rangle \langle n,0|
\end{equation}
we have:
\begin{eqnarray}\label{eq:dSqu-c}
    S_f-S_i&=&(1-e^{-\beta})\sum_{n} e^{-\beta n} \times  \\
& &[\left\langle \ln (m+1/2) \right\rangle_n-\ln (n+1/2)] \geq 0
\nonumber
\end{eqnarray}
\begin{figure}
\includegraphics[width=7 cm]{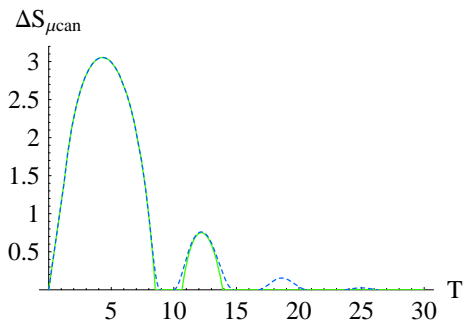}
  \caption{Change in microcanonical expectation of entropy for the force in Eq. (\ref{eq:f}) with $L=6$ , $n=2$, as a function of switching time $T$.
Solid line: classical case, Eq. (\ref{eq:dScl-mc}), with
$\Phi=n+1/2=5/2$. Dashed line: quantum case, Eq.
(\ref{eq:dSqu-mc}), the summation is truncated at $m=1000$.}
\label{fig:fig2}
\end{figure}
\begin{figure}
\includegraphics[width=7 cm]{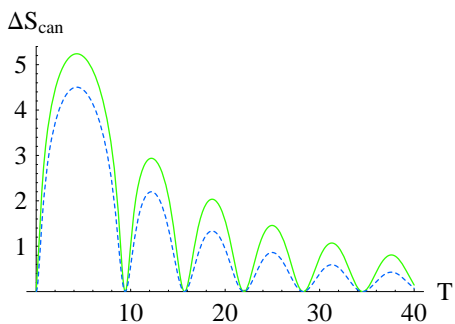}
  \caption{Change in canonical expectation of entropy for the force in Eq. (\ref{eq:f}) with $L=6$, $\beta=2$, as a function of switching time $T$.
The canonical distribution is truncated at $n=100$. Solid line:
classical case, Eq. (\ref{eq:dScl-mc}), with $\Phi=n+1/2=5/2$.
Dashed line: quantum case, Eq. (\ref{eq:dSqu-mc}), the summation
is truncated at $m=1000$.} \label{fig:fig3}
\end{figure}
Let us now study the actual change in expectation value of the
entropy for a specific shape of $f(t)$. Let
\begin{eqnarray}\label{eq:f}
    f(t)&=& L \sin\left(\frac{\pi t}{T}\right) \quad t \in [0,T]; \quad 0 \text{
    otherwise}\\
    W(T)&=& \frac{L^2 \pi ^2 T^2 (1+\cos T)}{\left(\pi
^2-T^2\right)^2}
\end{eqnarray}

Figure \ref{fig:fig2} shows the quantum and classical
microcanonical entropy change resulting from the force
(\ref{eq:f}), as a function of switching time $T$. Note that as
expected the change in entropy goes to zero in the adiabatic limit
$T\rightarrow \infty$. Note also a curious fact. The function
$W(T)$ is bounded from above $W(T)\leq c L^2$. Thus, for a given
initial energy $\Phi$ and amplitude $L$, if $L$ is not
sufficiently large, the change in microcanonical entropy remains
zero for all values of switching time $T$! This is evident in the
classical case for which one has $L\leq \sqrt{\Phi/c} \Rightarrow
W(T)\leq \Phi \Rightarrow \left\langle \ln\Theta
\right\rangle_\Phi = \ln \Phi, \forall T $ .

Figure \ref{fig:fig3} shows the quantum and classical canonical
entropy change. Note that, in all cases, in accordance with Eq.
(\ref{eq:Entropy-Increase}), the entropy change is
\emph{non-negative}.

The Quantum Mechanical Principle of Entropy Increase has been
established here by replacing von-Neumann entropy with a new
quantum entropy operator which is the natural quantum version of
the classical microcanonical entropy. The unitary character of the
quantum time evolution is responsible for keeping the former
always constant and making the latter grow. The validity of the
Quantum Mechanical Principle of Entropy Increase has been checked
on the driven harmonic oscillator. Interestingly, we have noticed
that the principle is satisfied in this example even in the
microcanonical ensemble, thus showing that the requirement of
decreasing probabilities is sufficient but not necessary. Further
investigations will aim at establishing under what general
conditions the principle is satisfied in the microcanonical case.
Fundamental implications, equivalence with the Minimal Work
Principle, and the mirror-image Principle of Entropy Decrease
(valid in the negative temperature scenario), have been discussed
elsewhere \cite{Campisi08,CampisiDiss}. These results contribute
to our understanding of \emph{quantum thermodynamics}. The Quantum
Mechanical Principle of Entropy Increase could be tested
experimentally, for example, with single molecule pulling
experiments.

An award from the Texas Section of the American Physical Society
for the presentation of this work at their Fall '07 meeting is
gratefully acknowledged.


\end{document}